\documentclass[a4paper,fleqn,usenatbib]{mnras}

\usepackage{newtxtext,newtxmath}

\usepackage[T1]{fontenc}
\usepackage{ae,aecompl}

\usepackage{graphicx}	
\usepackage{amsmath}	
\usepackage{amssymb}	

\def\teff{$T\rm_{eff}$}
\def\kms{$\mathrm{km\, s^{-1}}$}

\newcommand{\glog}{\ensuremath{\log g}}
\newcommand{\mygi}{MyGIsFOS}

\title[NLTE Copper computations]{Galactic evolution of Copper in the light of  NLTE computations\
\thanks{Based on observations taken at ESO, programmes 165.N-0276, 
65.L-0165,
65.L-0507,
266.D-5655,
66.B-0378,
67.D-0439,
68.D-0546,
70.D-0009,
71.B-0529,
072.B-0585,
078.D-0643 and HST programme
GO-14161}
}

\author[S. Andrievsky et al.]{
S. Andrievsky$^{1,2}$\thanks{E-mail: andrievskii@ukr.net},
P. Bonifacio$^{2}$,
E. Caffau$^{2}$,
S.~Korotin,$^{1,3}$,
M. Spite$^{2}$,
F. Spite$^{2}$,
\newauthor
L. Sbordone$^{4}$
and A.V. Zhukova$^{3}$
\\
$^{1}$Department of Astronomy and Astronomical Observatory, Odessa National University, Isaac Newton Institute of Chile, Odessa Branch,\\ 
Shevchenko Park, 65014, Odessa, Ukraine\\
$^{2}$GEPI, Observatoire de Paris, PSL Research University, CNRS, Place Jules Janssen, 92195 Meudon, France\\
$^{3}$Crimean Astrophysical Observatory, Nauchny 298409, Republic of Crimea\\
$^{4}$European Southern Observatory, Alonso de Cordova 3107, Vitacura, Santiago, Chile
}

\date{Accepted XXX. Received YYY; in original form ZZZ}

\pubyear{2015}

\begin{document}
\label{firstpage}
\pagerange{\pageref{firstpage}--\pageref{lastpage}}
\maketitle

\begin{abstract}
We have developed a model atom for Cu with which we perform statistical
equilibrium computations that allow us to compute the line formation of
\ion{Cu}{i} lines in stellar atmospheres without assuming Local Thermodynamic Equilibrium (LTE).
We validate this model atom by reproducing the observed line profiles of the Sun, Procyon 
and eleven metal-poor stars. Our sample of stars includes both dwarfs and giants.
Over a wide range of stellar parameters we obtain excellent
agreement among different \ion{Cu}{i} lines.
The eleven metal-poor stars have iron abundances in the range $-4.2 \le \rm [Fe/H] \le -1.4$,
the weighted mean of the [Cu/Fe] ratios is $-0.22$\,dex, with a scatter of $-0.15$\,dex. 
This is very different from the results from LTE analysis 
(the difference between NLTE and LTE abundances reaches 1 dex)
and in spite of the small size of our sample it prompts for a revision of the Galactic evolution of Cu.
\end{abstract}




\section{Introduction}

Copper is an odd element that can be formed through 
several  nucleosynthetic processes \citep{Bisterzo04},
the relevant importance of the various processes is debatable.
Up to now there has been a general consensus that, 
observationally, the [Cu/Fe] ratio decreases with decreasing
metallicity 
(e.g. \citealt{Cohen80,Sneden91,Mishenina02,Simmerer,Bihain},
see \citealt{Bonifacio10} for a concise summary of the 
observations). 
At very low metallicity the only lines that are strong
enough to be measured on ground-based spectra 
are the UV \ion{Cu}{i} resonant doublet at 3247\,\AA\ and 3273\,\AA.
\citet{Bihain} and \citet{Bonifacio10} pointed out
that, when computed under the assumptions of Local Thermodynamic
Equilibrium (LTE), these lines are discrepant with the optical lines. 
\citet{2014ApJ...791...32R} could measure vacuum UV \ion{Cu}{ii}
lines in two metal-poor stars, on spectra taken with the GHRS and  STIS
spectrographs on the Hubble Space Telescope and found that they
were strongly discrepant with the UV \ion{Cu}{i} resonant lines.
Collectively these observations prompt to consider the line formation
of \ion{Cu}{i} lines relaxing the hypothesis of Local Thermodynic Equilibrium. 
Such an investigation was conducted by \citet{Shi14}, however they 
only considered the optical lines and not the UV resonant doublet. 
The model atom developed by \citet{Shi14} was used by \citet{Yan15}
to investigate a sample of stars covering the range in metallicity 
--1.9 to --0.2. In this investigation they confirmed the decrease of
the [Cu/Fe] ratios with decreasing metallicity, albeit with higher
[Cu/Fe] ratios at the lowest metallicity, with respect to the previous LTE
analysis.  
In a subsequent investigation \citet{Yan16} claimed that their
NLTE analysis confirmed also the 
differences in Cu abundances between the high-$\alpha$
and low-$\alpha$ populations found by \citet{2011A&A...530A..15N}
in the metallicity range --1.5 to --0.4. 

The main purpose of our investigation is to see whether a NLTE approach is 
capable of reconciling the Cu abundances derived from the 
UV resonant doublet and the optical lines in metal-poor stars. 
For this purpose we developed a Cu model atom, and after a first
validation on the spectra of the Sun and Procyon we applied it
to a small sample of metal-poor stars covering the range
of --4.2 to --1.4 \relax in [Fe/H]. When possible we complemented
the ground-based spectra with vacuum UV spectra observed
with Hubble Space Telescope.



\section{Copper Model atom}
\begin{figure}
	\includegraphics[width=\columnwidth]{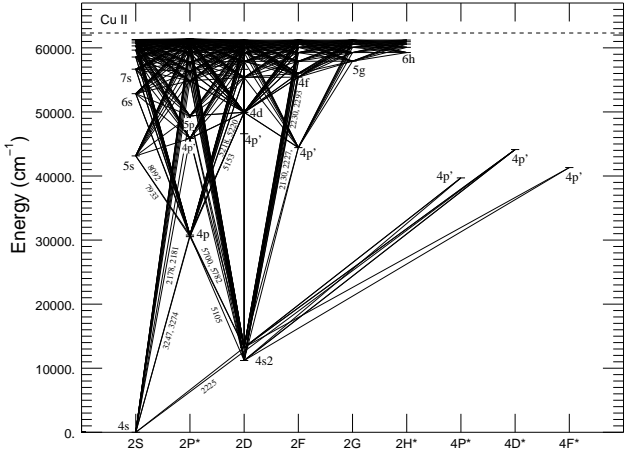}
    \caption{The Grotrian diagram of our Cu model atom.}
    \label{grotrian}
\end{figure}

To build a model  atom of copper we used 130 levels (116 levels of \ion{Cu}{i} and 14 levels of \ion{Cu}{ii}). 
Parameters of atomic levels were taken from \citet{Liu2014}, \citet{1990JPCRD..19..527S} 
and the NIST database. Each LS multiplet was considered as a single term. 
The fine structure was taken into account for the following levels: 4s$^{2}$~2D, 4p$^{2}$~P$^{\rm o}$ and 4d~2D, 
that are tightly coupled 
with the most important transitions in the \ion{Cu}{i} atom (it should be noted that the fine splitting in this 
case is very important: for instance the splitting of 4s$^{2}$~2D is larger than 0.25\,eV).
The highest energy level in our \ion{Cu}{i} model 
for which we accounted for the radiative and collisional transitions has a ionization potential 0.11\,eV that 
corresponds to an electronic temperature of about 1300\,K. 
Thus, we can conclude that our model describes the connection between excited levels and continuum quite accurately. 
The adopted Grotrian diagram of or Cu atomic model is shown in Fig.\ref{grotrian}.

We have considered radiative transitions between the first 59 levels of 
\ion{Cu}{i} and ground level of \ion{Cu}{ii}. 
Populations of the other levels were taken into account only in the solution of the particle number equation. 
All 486 bound-bound transitions were considered in detail. For each transition we accounted for Stark and van der Waals effects, 
as well as for the influence of the  microturbulent velocity. To adequately reproduce the radiative rates of the 
several significant UV transitions from 4s~2S and 4s$^{2}$~2D levels we calculated the corresponding profiles for 
100 to 150 frequency points. For the rest of the line profiles we used only 30 frequency points.
Photoionization cross-sections were taken from \citet{Liu2014}. For all the bound-bound transitions 
the oscillator strengths were compared to those from the NIST database, and  if $\log$~gf values were absent there, 
we used the values provided by theoretical calculations of \citet{Liu2014}.

Collisional ionization rates were calculated basing on the Seaton's  formula \citep{Seaton62} with threshold 
photoionization cross-section value from \citet{Liu2014}. Collisional excitations by electrons were calculated with 
the help of van Regemorter's formula (\citealt{vanRegemorter}). 
Collisional rates for the forbidden transitions can be found with Allen's 
formula (\citealt{Allen73})  with an effective collisional strength of 1. 
For the 30 transitions between the low excited levels we used the collisional rates 
calculated by M. O'Mullane, and made available
through the Atomic Data and Analysis Structure \citep[ADAS,][]{ADAS}. 
Inelastic collisions with hydrogen atoms were described with the Drawin's formula \citep{Drawin68,Drawin69} 
adapted for astrophysical use by \citet{SH84}.

Atomic level populations were determined using the MULTI code of \citet{Car86} 
with modifications as given in \citet{Kor99}. 
MULTI calculates the line profile for each line considered in detail. The line 
profile computed assuming either LTE or NLTE depends upon many parameters: the 
effective temperature of the model, the surface gravity, the microturbulent 
velocity, and the line damping as well as the populations in the appropriate 
levels. Proper comparison of observed and computed profiles in many cases 
requires a multielement synthesis to take into account possible blending lines 
of other species.  For this process, we fold the NLTE (MULTI) calculations, 
specifically the departure coefficients, into the LTE synthetic spectrum code 
SYNTHV (\citealt{Tsym96}) that enables us to calculate the NLTE source function 
for copper lines.  These calculations included all spectral lines from the VALD 
database (\citealt{2015PhyS...90e4005R}) in a region of interest. The LTE approach was applied for lines other 
than the \ion{Cu}{i} lines. Abundances of corresponding elements were adopted 
in accordance with the [Fe/H] value for each star.

\ion{Cu}{i} has several lines that are suitable for abundance determination in the visual range: 
5105\,\AA, 5153\,\AA, 5218\,\AA, 5220\,\AA, 5700\,\AA, 5782\,\AA\,
(oscillator strengths are from \citealt{1968ZA.....69..180K}), 
and in IR range: 7933\,\AA, 8092\,\AA\,
(oscillator strengths are from \citealt{1988PhRvA..38.1702C}). 
Two resonance UV lines 3247\,\AA~and 3273\,\AA~ are too strong in the solar-metallicity stars. 
Nevertheless, these lines can be important as a copper abundance indicator in the metal-poor stars \citep{Bihain,Bonifacio10}. 
The $\log$~gf value for 5782 \AA~line is $-1.78$ ( $\pm 12$\%) in \citet{1968ZA.....69..180K}, and $-1.91$ in 
\citet{Liu2014}. From our solar spectrum analysis we obtained $\log$~gf = $-1.83$. 
All the line parameters we used are given in Table\,\ref{atomdata}. 
We adopted the following isotopic ratio ${}^{63}$Cu to $^{65}$Cu =
0.69/0.31 (recommended by \citealt{2015A&A...573A..27G}). 
For most of the lines we used the hyperfine structure ({\it hfs}) component ratios and wavelength shifts from \citet{Shi14}. 
As a control, we also performed the synthesis of the \ion{Cu}{i} line profiles in the 
solar spectrum using the corresponding data from Kurucz's list 
\citep{2014dapb.book...63K,2011CaJPh..89..417K,2005MSAIS...8...86K}. 
We did not obtain any significant differences. For the better reproduction of the profiles of the resonance doublet 
3247\,\AA, and 3273\,\AA\ and the high excitation line 
5105\,\AA\ line in the solar spectrum we used the 
{\it hfs} component shifts from the magnetic dipole splitting constants A(J) 
and electric quadrupole splitting constants B(J) from \citet{1979OptCo..31...28G} and \citet{1993AcSpe..48.1259H}.
Components with wavelength differences less than 0.001\,\AA\
were combined into one component. Results are given in Table\,\ref{atomdata}.

\begin{figure*}
\centering
	\includegraphics[width=18cm,clip=true]{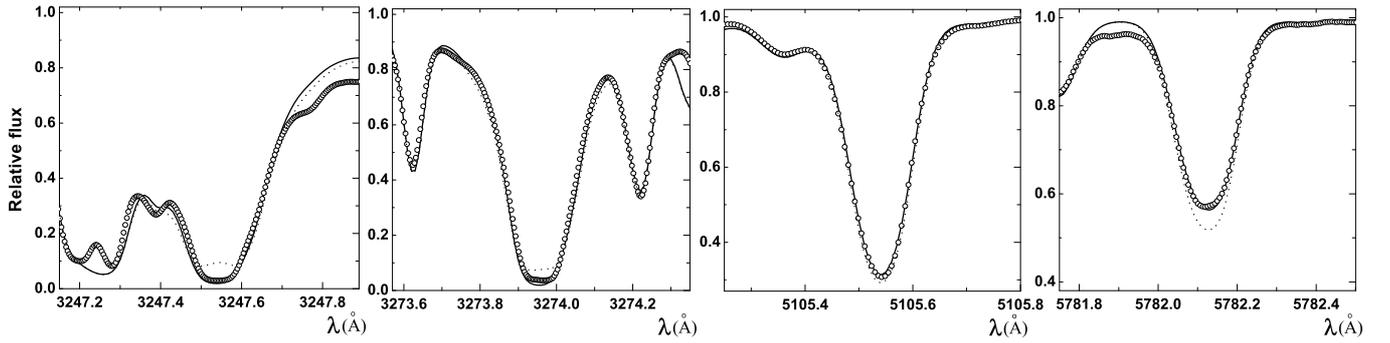}
    \caption{Several \ion{Cu}{i} lines in the observed flux spectrum of the Sun \citep{1984sfat.book.....K}, open circles,
compared to our NLTE synthetic spectrum (solid) line and LTE synthetic spectrum (dotted line). }
    \label{solcu}
\end{figure*}

\begin{figure*}
\centering
	\includegraphics[width=18cm,clip=true]{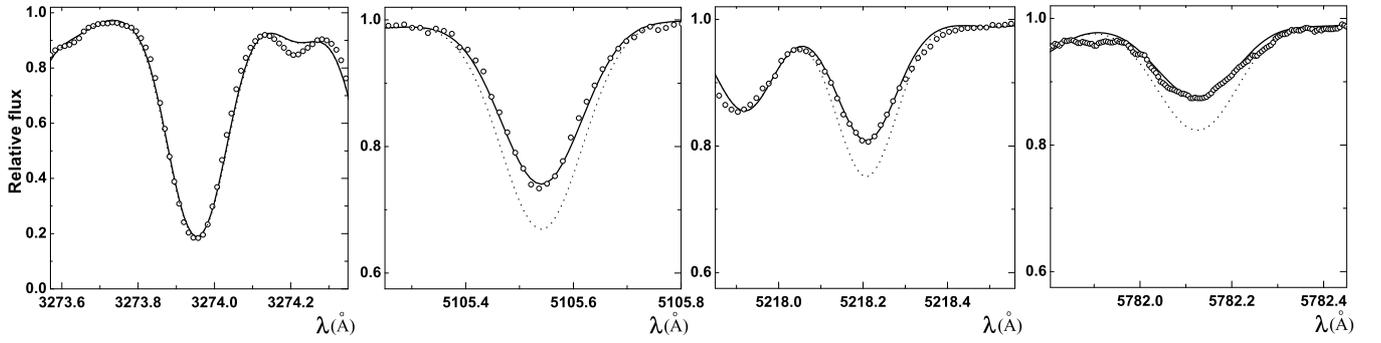}
    \caption{Several \ion{Cu}{i} lines in the observed spectrum of the Procyon, open circles,
compared to our NLTE synthetic spectrum (solid) line and LTE synthetic spectrum (dotted line). }
    \label{cu_procyon}
\end{figure*}

Since the strong \ion{Cu}{i} lines are formed in the upper atmosphere layers we 
applied the combination of the ATLAS solar model atmosphere of 
\citet{CK03} 
with a chromosphere from the VAL-C model  
\citet{Vernazza}
and with corresponding distribution of the microturbulent velocity. 
Observed line profiles were taken from the solar flux spectrum 
\citep{1984sfat.book.....K}.
All the line profiles were calculated assuming the  Cu meteoritic abundance 
(A(Cu) = log (Cu/H) + 12 = 4.25, \citealt{Lodders}). To reproduce the copper lines in
the  solar spectrum we used the collisional rates with atomic hydrogen atoms 
without any correcting factor. 
If one uses the value adopted by \citet{Shi14}, then the NLTE effects lead to a 
significant unbalance of the calculated line intensities and
this implies an increase the copper abundance in solar atmosphere. 
We show the calculated and observed profiles of some lines in the solar spectrum in Fig.\,\ref{solcu}. 
All the investigated \ion{Cu}{i} lines are weakened by the NLTE effects. 
Moreover, the line cores cannot be correctly described within the LTE approximation.
As a further verification of our \ion{Cu}{i} model atom
we studied the spectra of Procyon.
We obtained  an excellent agreement between observed and calculated 
\ion{Cu}{i} profiles  adopting a  copper abundance A(Cu) = 4.30 $\pm 0.03$. 
(see Fig.\,\ref{cu_procyon}).
%
\begin{table}
\centering
\caption{Programme stars and adopted stellar parameters.}
\renewcommand{\tabcolsep}{1pt}
\tabskip=0pt

\label{stars}
\begin{tabular}{lcccccc} 
\hline
Name & \teff & \glog & $\xi$ & [Fe/H] & Ref. & A(Cu)  \\
     & K     & g in cgs & \kms  &  dex   &  & dex \\
\hline
HD\,111721           & 5095 & 2.64  &  1.4  & $-1.40$ & 1 & $ 2.87 \pm 0.05$ \\ 
HD\,94028             & 5970 & 4.33  &  1.3  & $-1.47$ & 2 & $ 2.60 \pm 0.03$ \\ 
Cl*\,NGC\,6752\,YGN\,30 & 4943 & 2.42  &  1.3  & $-1.62$ & 3 & $ 2.35 \pm 0.10$ \\ 
HD\,9051             & 4925 & 1.95  &  1.8  & $-1.78$ & 1 & $ 2.41 \pm 0.04$ \\ 
HD\,84937             & 6300 & 4.00  &  1.3  & $-2.25$ & 4 & $ 1.80 \pm 0.02$  \\
HD\,128279            & 5040 & 2.22  &  1.4  & $-2.45$ & 1 & $ 1.50 \pm 0.06$ \\ 
HD\,140283           & 5750 & 3.70  &  1.4  & $-2.59$ & 5 & $ 1.40 \pm 0.02$ \\ 
HD\,122563            & 4600 & 1.10  &  2.0  & $-2.82$ & 6 & $ 1.18 \pm 0.03$ \\ 
CS\,31082-001        & 4825 & 1.50  &  1.8  & $-2.91$ & 6 & $ 1.35 \pm 0.10$ \\ 
HD\,200654           & 5007 & 2.21  &  1.0  & $-3.16$ & 1 & $ 0.91 \pm 0.05 $ \\ 
CD\,--38$^\circ$\,245 & 4800 & 1.50  &  2.2  & $-4.19$ & 6 & $ 0.04 \pm 0.10$ \\ 
\hline
\multicolumn{7}{l}{References for the atmospheric parameters}\\
\multicolumn{7}{l}{1. This paper}\\
\multicolumn{7}{l}{2. \citet{Sitnova}}\\
\multicolumn{7}{l}{3. \citet{Yong}}\\
\multicolumn{7}{l}{4. \citet{Spite17}}\\
\multicolumn{7}{l}{5. \citet{2015A&A...584A..86S}}\\
\multicolumn{7}{l}{6. \citet{FirstStarsV}}\\
\hline
\end{tabular}
\end{table}

\section{Copper abundances in stars of different metallicity}

In order to gain some insight in the Galactic evolution
of Cu we applied our model atom and NLTE analysis to a set of stars
for which we could find good quality visual and UV spectra in the
ESO archive and which span a large range in metallicity. 
We requested that for each star the spectrum of the UV doublet
be available, this is covered for all stars by observations obtained
with UVES.  
For two stars (HD\,84937 and HD\,140283) we could also use
HST STIS spectra, that gave us access to the vacuum ultra-violet. 
For  HD\,140283 we also made use of HARPS spectra.
For HD\,111721, HD\,9051 and HD\,128279 we determined the
atmospheric parameters from the analysis of the visible spectra.
We used the \mygi\ code \citep{mygisfos} and 
the grids of synthetic spectra used by the \mygi\ code
in the Gaia-ESO survey \citep[see][]{2014A&A...570A.122S,duffau}.
The temperatures were derived from the iron excitation equilibrium
and the surface gravities from the iron ionisation equilibrium.
For the other stars we adopted atmospheric parameters from the literature.
The names, adopted stellar parameters and NLTE Cu abundances
are listed in Table\,\ref{stars}.

For each star we computed a 1D LTE model atmosphere
using version 9 of the ATLAS code \citet{1993KurCD..13.....K,2005MSAIS...8...14K}
in the Linux version detailed by \citet{2004MSAIS...5...93S,2005MSAIS...8...61S}
and the Opacity Distribution Functions of \citep{CK03} with microturbulence
of 1\,\kms.

\begin{table}
\caption{Atomic data for the lines \ion{Cu}{i}.}
\renewcommand{\tabcolsep}{3pt}
\tabskip=0pt
\label{atomdata}
\begin{tabular}{cccccccc}
\hline
\hline
$\lambda$& log gf  & log gf      & Ref&$\lambda$ & log gf  & log gf      & Ref\\ 
(\AA)    & {\it hfs}&  line       &    &  (\AA)   & {\it hfs}& line        &    \\ 
\hline                                                                           
2165.096 &         &-0.840       &  1 & 5700.143 &   -4.298&     -2.58	 &  3 \\ 
2178.949 &         &-0.586       &  1 & 5700.159 &   -4.094&     	 &    \\ 
2181.722 &         &-0.741       &  1 & 5700.164 &   -4.696&             &    \\ 
2199.586 &         & 0.447       &  1 & 5700.166 &   -4.298&             &    \\ 
2199.754 &         & 0.340       &  1 & 5700.173 &   -3.948&             &    \\ 
2214.583 &         & 0.108       &  1 & 5700.188 &   -3.744&             &    \\ 
2225.705 &         &-1.205       &  1 & 5700.193 &   -4.346&             &    \\ 
2227.776 &         & 0.460       &  1 & 5700.194 &   -4.152&             &    \\ 
2230.086 &         & 0.642       &  1 & 5700.195 &   -3.948&             &    \\ 
2293.844 &         &-0.115       &  1 & 5700.200 &   -3.997&             &    \\ 
         &         &             &    & 5700.205 &   -4.094&             &    \\ 
3247.511 &   -1.379&     -0.05	 &  2 & 5700.221 &   -3.802&             &    \\ 
3247.513 &   -1.028&     	 &    & 5700.227 &   -3.647&             &    \\ 
3247.515 &   -1.379&     	 &    & 5700.231 &   -3.744&             &    \\ 
3247.517 &   -0.957&     	 &    & 5700.260 &   -3.550&             &    \\ 
3247.520 &   -1.426&     	 &    & 5700.266 &   -4.152&             &    \\ 
3247.555 &   -0.288&     	 &    & 5700.279 &   -3.198&             &    \\ 
3247.558 &   -1.567&     	 &    & 5700.285 &   -3.802&             &    \\ 
	 &   	   &             &    & 	 &   	   &             &    \\ 
3273.927 &   -1.375&     -0.35	 &  2 & 5782.034 &   -3.544&     -1.83	 &  4 \\ 
3273.929 &   -1.024&     	 &    & 5782.042 &   -3.845&     	 &    \\ 
3273.931 &   -2.074&     	 &    & 5782.054 &   -3.146&     	 &    \\ 
3273.933 &   -1.723&     	 &    & 5782.064 &   -3.196&     	 &    \\ 
3273.971 &   -1.024&     	 &    & 5782.073 &   -3.497&     	 &    \\ 
3273.972 &   -1.375&     	 &    & 5782.084 &   -2.798&     	 &    \\ 
3273.975 &   -1.024&     	 &    & 5782.085 &   -3.146&     	 &    \\ 
3273.976 &   -1.375&     	 &    & 5782.097 &   -3.146&     	 &    \\ 
	 &   	   &             &    & 5782.113 &   -2.798&     	 &    \\ 
5105.504 &   -3.720&     -1.51	 &  2 & 5782.124 &   -2.798&     	 &    \\ 
5105.513 &   -2.766&     	 &    & 5782.153 &   -2.698&     	 &    \\ 
5105.516 &   -2.813&     	 &    & 5782.172 &   -2.350&     	 &    \\ 
5105.517 &   -3.090&     	 &    & 	 &   	   &             &    \\ 
5105.521 &   -2.720&     	 &    & 7933.125 &   -0.420&             &  5 \\ 
5105.526 &   -2.398&     	 &    & 	 &   	   &             &    \\ 
5105.536 &   -2.051&     	 &    & 8092.603 &   -2.179&     -0.16	 &  5 \\ 
5105.563 &   -1.942&     	 &    & 8092.604 &   -1.831&     	 &    \\ 
	 &   	   &             &    & 8092.612 &   -0.971&     	 &    \\ 
5153.232 &   -0.441&     -0.01	 &  2 & 8092.625 &   -0.523&     	 &    \\ 
5153.241 &   -0.219&     	 &    & 8092.638 &   -1.530&     	 &    \\ 
	 &   	   &             &    & 8092.639 &   -1.878&     	 &    \\ 
5218.200 &   -0.934&      0.27	 &  2 & 8092.642 &   -1.132&     	 &    \\ 
5218.202 &   -0.457&     	 &    & 8092.644 &   -1.480&     	 &    \\ 
5218.206 &   -0.236&     	 &    & 8092.651 &   -1.132&     	 &    \\ 
5218.211 &   -0.089&     	 &    & 8092.653 &   -1.480&     	 &    \\     
	 &   	   &             &    &&&&\\                                     
5220.071 &   -1.816&     -0.61	 &  2 &&&&\\                                     
5220.073 &   -1.338&     	 &    &&&&\\                                     
5220.077 &   -1.117&     	 &    &&&&\\                                     
5220.082 &   -1.131&     	 &    &&&&\\                                     
5220.083 &   -1.481&     	 &    &&&&\\                                     
\hline
\end{tabular}
Notes: References for log gf\\
1. VALD database \citet{2015PhyS...90e4005R}\\
2. \citet{1968ZA.....69..180K}\\
3. \citet{2014dapb.book...63K}\\
4. SUN\\
5. \citet{1988PhRvA..38.1702C}\\
\end{table}


\begin{figure*}
	\includegraphics[width=18cm,clip=true]{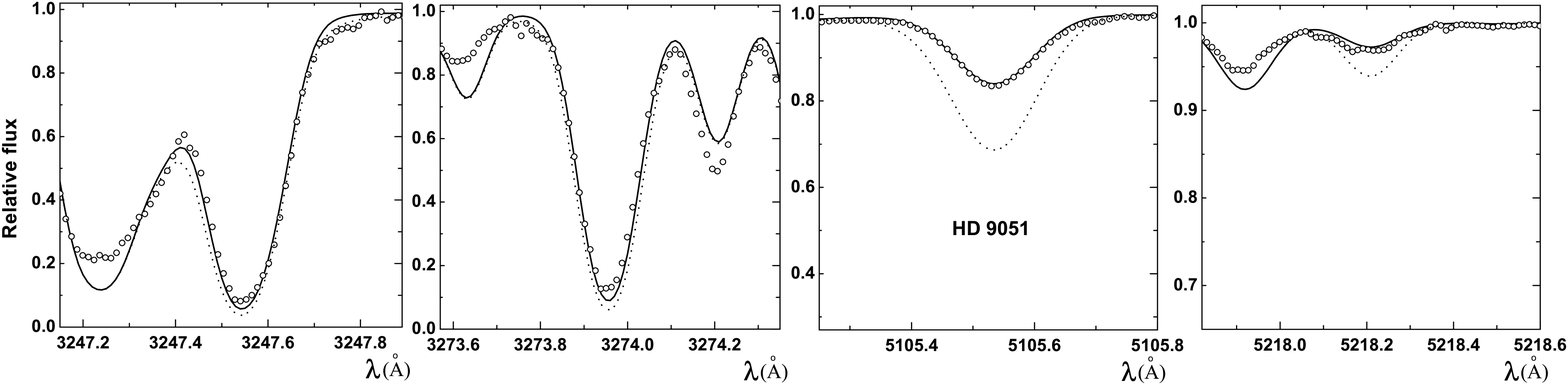}
	\includegraphics[width=18cm,clip=true]{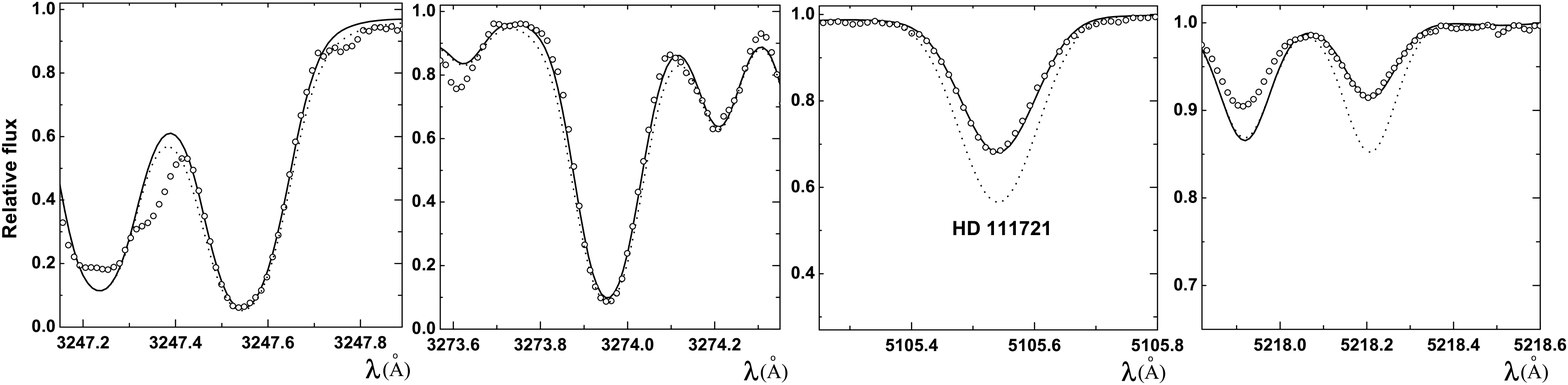}
    \caption{\ion{Cu}{i} lines in the observed  spectrum of the HD\,9051 (upper panel), HD\,11721 (lower panel), 
open circles,
compared to our NLTE synthetic spectrum (solid) line and LTE synthetic spectrum (dotted line). }
    \label{HD9051_HD111721}
\end{figure*}




%
\begin{figure*}
	\includegraphics[width=18cm,clip=true]{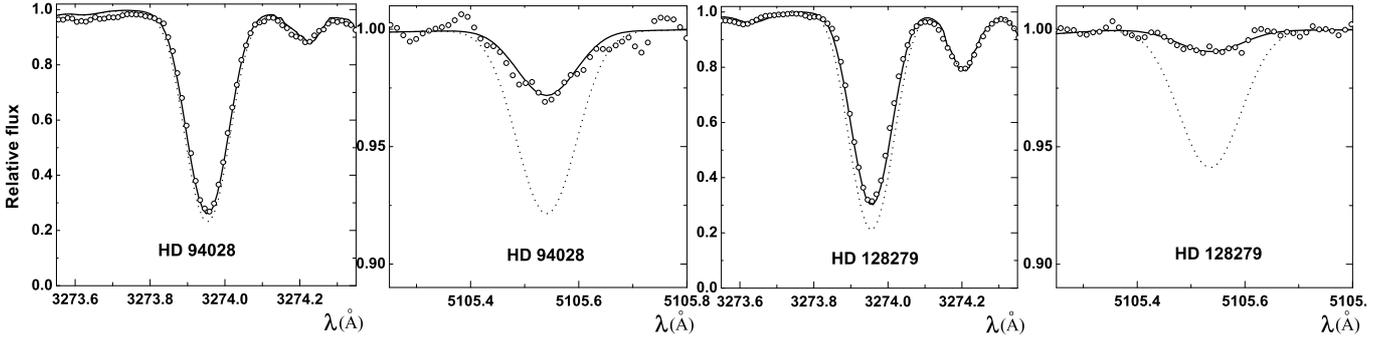}
    \caption{\ion{Cu}{i} lines in the observed  spectrum of the HD\,94028 (two panels on the left) 
and HD\,128279 (two panels on the right), open circles,
compared to our NLTE synthetic spectrum (solid) line and LTE synthetic spectrum (dotted line). }
    \label{HD94028_HD128279}
\end{figure*}

\begin{figure*}
	\includegraphics[width=18cm,clip=true]{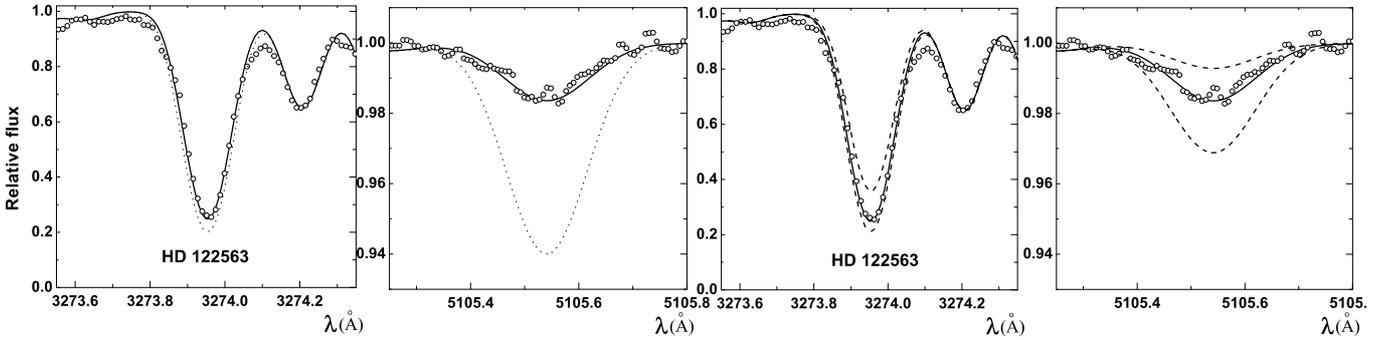}
    \caption{\ion{Cu}{i} lines in the observed  spectrum of the HD\,122563, open circles,
compared to our NLTE synthetic spectrum (solid) line and LTE synthetic spectrum (dotted line). 
	The two panels on the left show the best fitting NLTE profiles, the two panels
	on the right show NLTE profiles computed with a Cu abundance of $\pm 0.3$\,dex
	of the best fitting abundance.}
    \label{HD122563}
\end{figure*}

\begin{figure*}
	\includegraphics[width=18cm,clip=true]{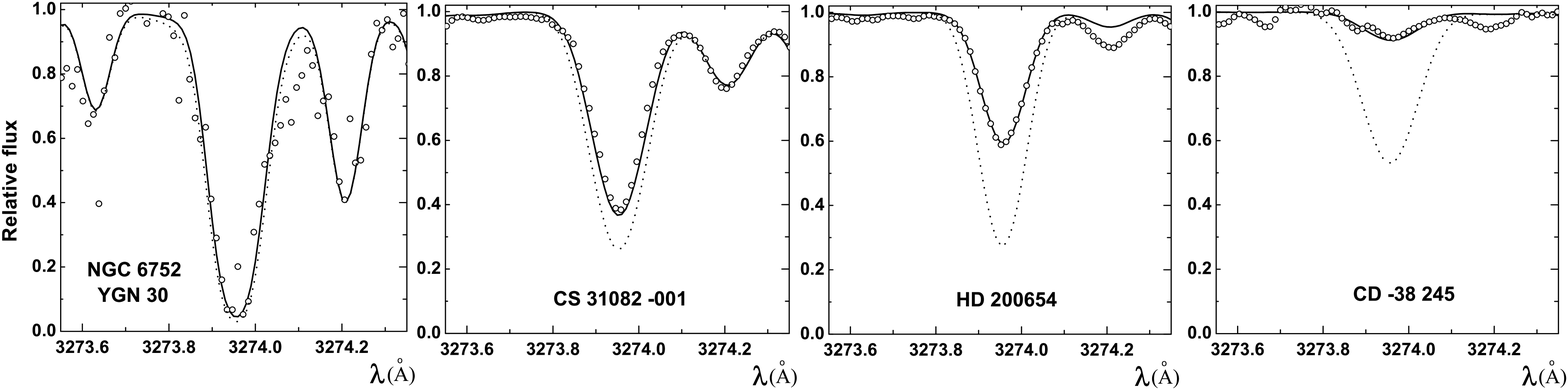}
    \caption{The \ion{Cu}{i} 3273\,\AA\  line in 
Cl* NGC 6752 YGN 30, CS\,31082-001, HD\,200654 and CD\, $-38^\circ 245$, open circles,
compared to our NLTE synthetic spectrum (solid) line and LTE synthetic spectrum (dotted line). }
    \label{Others}
\end{figure*}

For all of the programme stars we obtain a good consistency
of the abundance implied by the different lines, as shown in Fig.\,\ref{HD9051_HD111721} and \ref{HD94028_HD128279}. 
In particular for the metal-poor giants we always obtain the consistency
between the UV doublet and the high excitation  5105 \AA\ line.
In spectra of two program stars HD\,9051 and HD\,111721 the line 5218 \AA~ is 
also seen (Fig.\,\ref{HD9051_HD111721}). Its profile is well reproduced by our 
NLTE calculations. 
Accuracy of the profile fitting can be estimated, for instance, 
in Fig.\,\ref{HD122563} where we show LTE and NLTE profiles (two left panels), 
and NLTE profile variation with best copper abundance changed by $\pm 0.3$ dex 
(two right panels).
\citet{Bonifacio10} pointed out that this was never possible in LTE, even
when using 3D hydrodynamical simulations. 
In Fig.\,\ref{HD122563} and \ref{Others} we show that the NLTE profiles
reproduce well the observed profiles.

\begin{figure*}
	\includegraphics[width=18cm,clip=true]{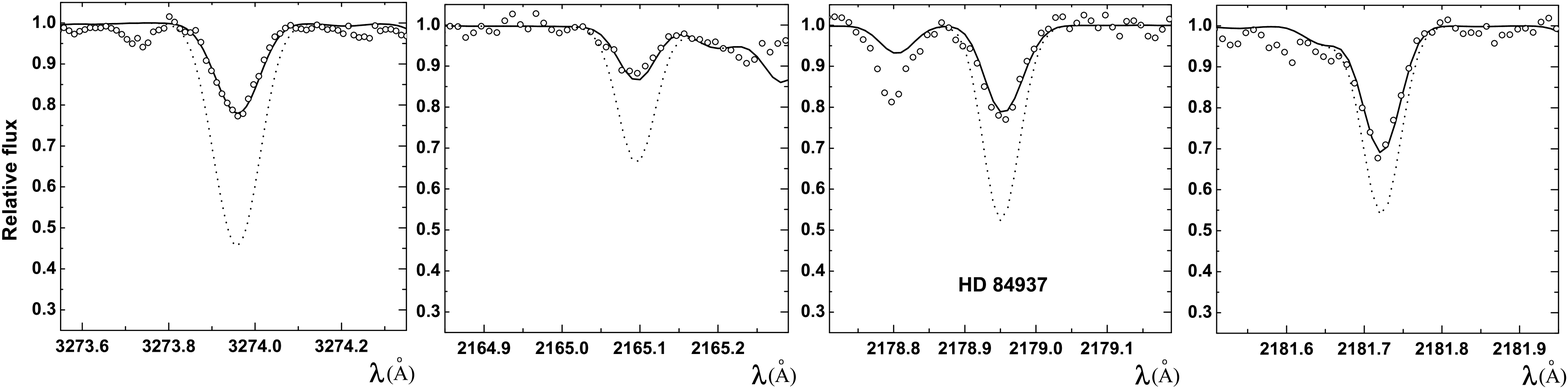}
	\includegraphics[width=18cm,clip=true]{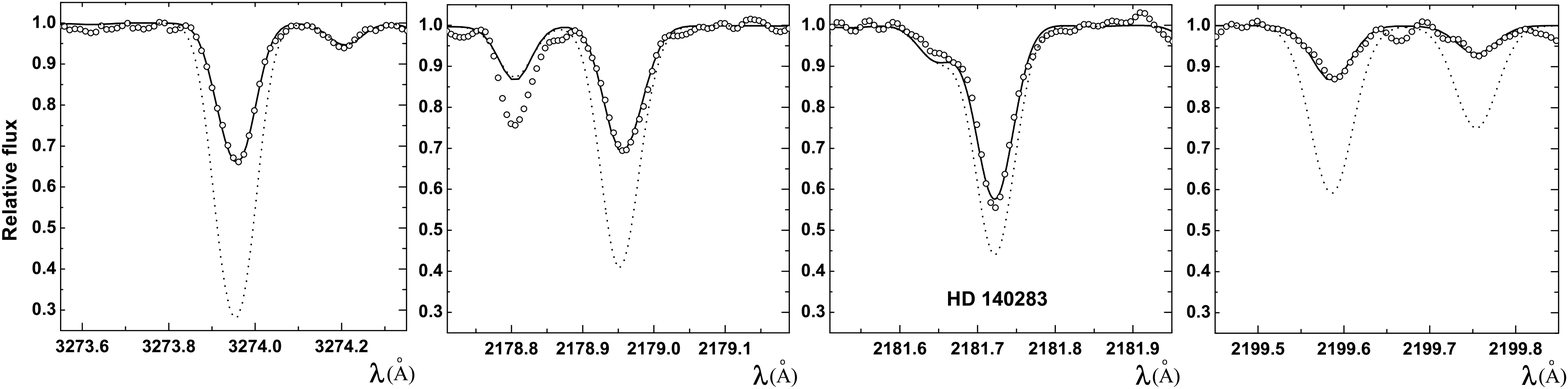}
    \caption{The  \ion{Cu}{i} 3273\,\AA, 2165\,\AA, 2178\,\AA, 2181\,\AA, and  2199\,\AA\ lines in the  spectrum of the HD\,84937 (upper panel) and 
HD\,140283 (lower panel), open circles,
compared to our NLTE synthetic spectrum (solid) line and LTE synthetic spectrum (dotted line). }
    \label{HD84937_HD140283}
\end{figure*}

We consider as a success of our NLTE model atom the fact that for
HD\,84937 and HD\,140283 we obtain consistency between 
the UV doublet 3242\,\AA\, and 3273\,\AA\, 
and the vacuum UV lines: 2165\,\AA, 2178\,\AA, 2181\,\AA, 2199\,\AA, 2214\,\AA, 2225\,\AA, 2227\,\AA,
and  2230\,\AA.
The 3273\,\AA, 2165\,\AA, 2178\,\AA, 2181\,\AA, and  2199\,\AA\ lines in HD\,84937 and HD\,140283 
are shown in Fig.\,\ref{HD84937_HD140283}.
For these two stars, we also derived the  copper abundance in LTE 
from four \ion{Cu}{ii} lines: 
2112.100\,\AA, 2126.044\,\AA, 2148.984\,\AA\, and 2247.003\,\AA. 
Our copper model atom is not designed to compute the NLTE
of the \ion{Cu}{ii} transitions. Since ionised
copper is the dominant species  in the atmospheres of our programme
stars we expect deviations from LTE to be small.
The parameters of the lines 
were taken from the VALD database (\citealt{2015PhyS...90e4005R}). We obtained $<$(Cu/H)$>$ = 1.71 $\pm$ 0.07 for 
HD\,84937 and $<$(Cu/H)$>$ = 1.31 $\pm$ 0.08 for HD\,140283. 
We consider the  
agreement satisfactory,  the difference with the 
NLTE-results for \ion{Cu}{i} is less 
than 0.10 dex. We stress that for \ion{Cu}{i} the difference between LTE and NLTE is 
of 0.70--0.80\,dex.

\begin{figure}
	\includegraphics[width=0.9\columnwidth,clip=true]{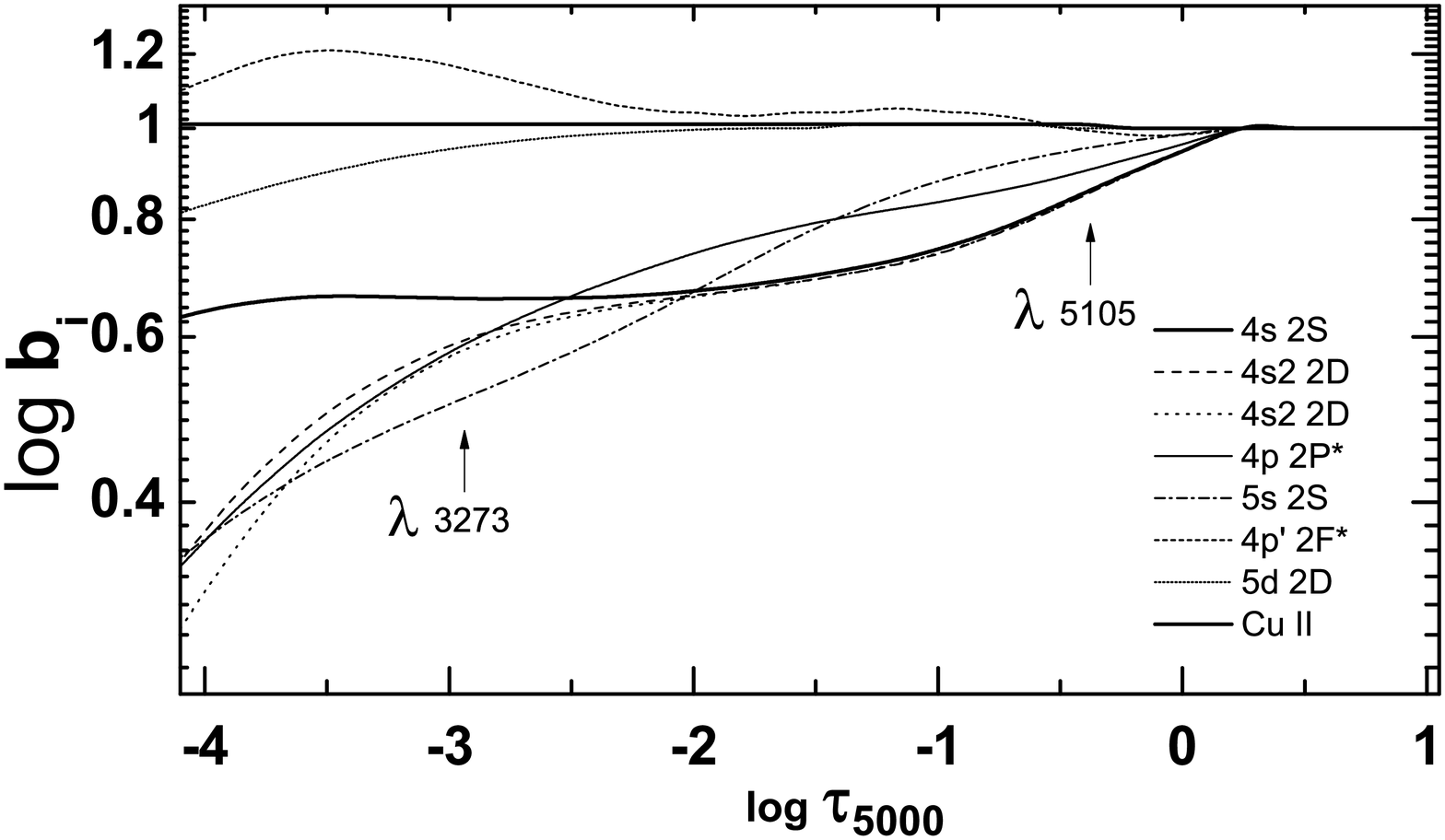}
	\includegraphics[width=0.9\columnwidth,clip=true]{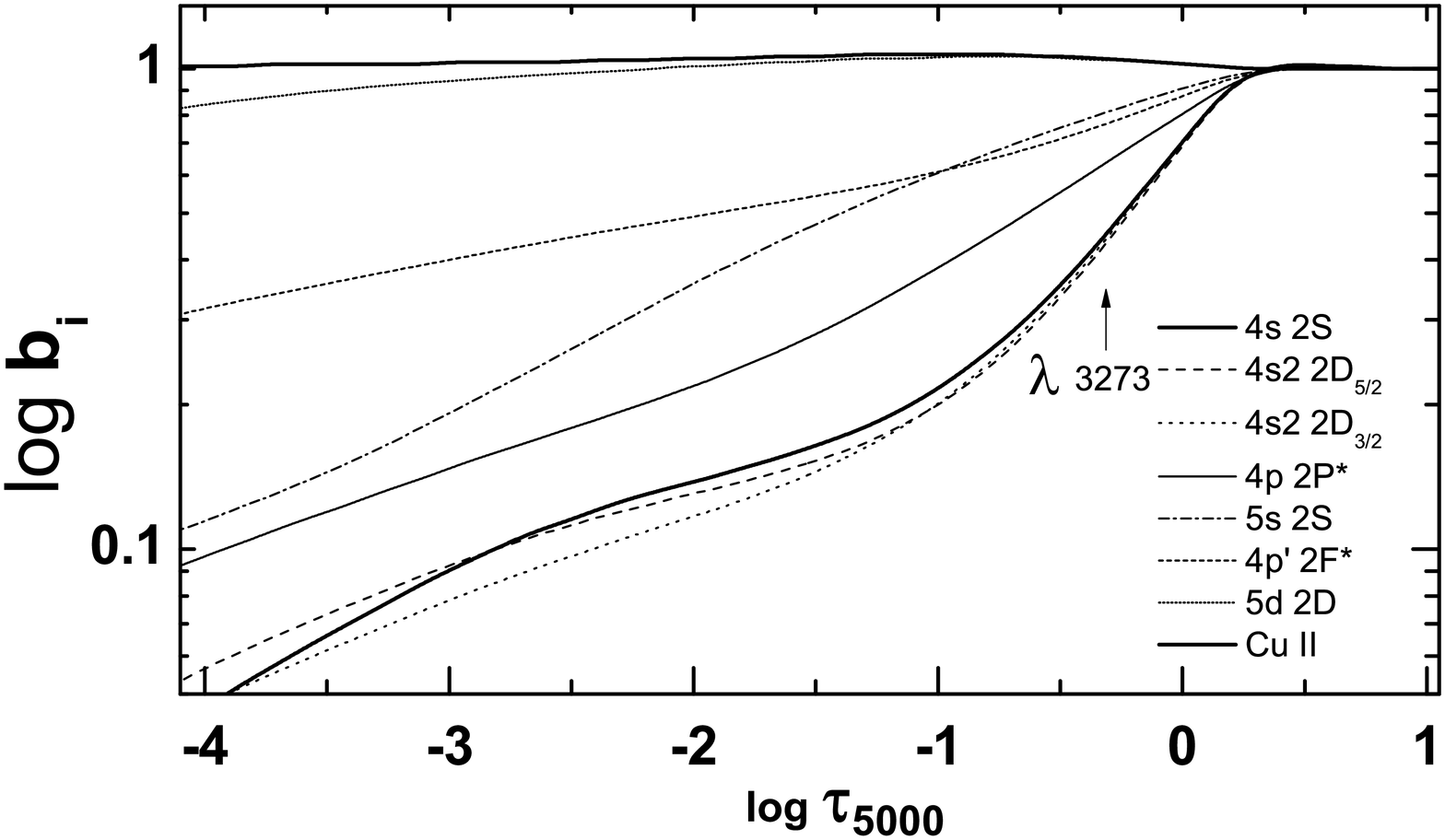}
    \caption{Computed departure coefficients as a function of optical depth in the atmosphere of Procyon (upper panel)
and HD\,140283 (lower panel).}
    \label{depcoef}
\end{figure}

As it was noted in 
\citet{Shi14} and  \citet{Yan15,Yan16} 
the deviations from LTE lead to the copper overionization 
already 
in the deep atmosphere layers. 
In Fig. \ref{depcoef} the distribution of departure coefficients 
($b_{\rm i} = \frac{n_{{\rm i}_{\rm NLTE}}}{n_{{\rm i}_{\rm LTE}}}$) 
are shown for Procyon (the star with solar metallicity) and the metal deficient star HD140283. 
As one can see, the deviations from LTE start to become important 
at the optical depth of about 1.
For the stars with solar metallicity the resonant doublet is practically not affected by the NLTE effects. 
These effects are gradually increasing as metallicity decreases, and reach 
large values (--0.7 dex for HD\,84937 and --0.75 dex for HD\,140283).
Subordinate copper lines show stronger NLTE effects even for the stars with solar metallicity, which are 
seen for example in  Procyon. 

As in the case of UV resonance doublet, the NLTE effects increase as metallicity decreases, 
but with different magnitude for different lines in some cases it can be as large as  1\,dex.
At the same time  it should be noted 
that in the spectra of metal-poor stars subordinate lines become too weak to be accurately measured. 
For the stars with \teff about 6000 K and with [Fe/H] of about --1.5 only the 
subordinate line at 5105\,\AA~ is available for analysis. In the spectra of 
cooler stars, the line 5782\,\AA~ can also be seen. These lines rapidly weaken 
with decreasing metallicity and completely disappear in the spectra of all 
stars at [Fe/H] of about --2.5.
The NLTE corrections significantly depend on 
atmospheric parameters (like \teff, \glog, $V_{\rm t}$, [Fe/H]), as well as on the copper abundance itself.

\section{Discussion}

Our sample of stars is very small and spans the metallicity range 
$-4.2\le \rm [Fe/H] \le -1.5$. We do not claim that we are in the
best possible position to discuss the Galactic evolution of copper,
yet a few results are already noticeable. 
The fact that our NLTE analysis derives consistent abundances
from different lines, including the strong vacuum UV lines, when available,
allows us to conclude that the UV resonant doublet at 3247\,\AA\ and 3273\,\AA\ is
a reliable indicator of the copper abundance. This is important because this
doublet is the only one available to measure the copper abundance in low
metallicity stars. We consider quite remarkable that we could measure
the doublet in CD\,--38$^\circ$\,245, that is, to our knowldedge, 
the measurement of Cu at lowest metallicity so far achieved. 
We also checked the  UVES spectra of HE\,0107-5240 ([Fe/H]$=-5.46$  \citealt{chris2004}),
but we could not detect any of the two lines. 
Based on the UVES spectra of CD\,--38$^\circ$\,245  the lines can be detected down to [Fe/H]=--4.8 
\relax in cool giants. 
Up to now the measurements of Cu at very low metallicity 
(\citealt{Bihain,Lai08,Cohen2008,2012ApJS..203...27R,2014ApJ...791...32R})
rely on the UV resonant lines. 
Our investigation shows that these measurements have to be revised to take
into account the NLTE effects on these lines.

\begin{figure}
	\includegraphics[width=0.9\columnwidth,clip=true]{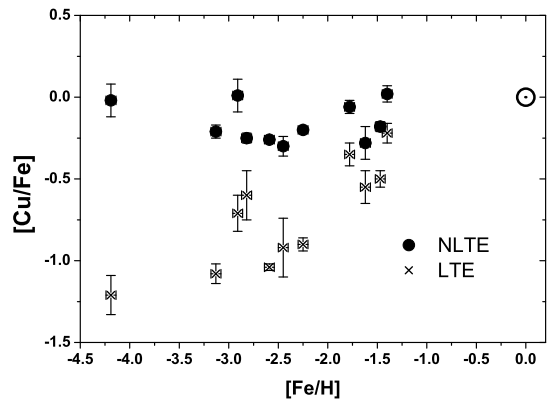}
    \caption{The Galactic evolution of Cu, as captured by our programme stars.
	Filled symbols are the NLTE results, $\times$ signs are the LTE results.}
    \label{cu_fe}
\end{figure}

In Fig.\,\ref{cu_fe} we show the run of [Cu/Fe] as a function of [Fe/H]
for our programme stars. We also show the solar value, for reference. 
The main result that is apparent is that there are no very low [Cu/Fe] values
at variance with what happens in the LTE analysis. 
Our sample is more metal-poor than the sample of \citet{Yan15}, and we have only
one star, HD\,94028 \relax in common. The adopted stellar parameters for this star
are very close, yet we provide [Cu/Fe]$=-0.18\pm 0.03$
while \citet{Yan15} provide $-0.32\pm 0.06$. 
Their abundance is based only on the 5105\,\AA\ line while ours relies
also on the 2 UV resonance lines. 
From the  5105\,\AA\ line we derive [Cu/Fe]$=-0.14$ and from the two
resonance lines --0.22. 
Our NLTE correction for the  5105\,\AA\ line is +0.48\,dex, while 
\citet{Yan16} have a correction that is only +0.17\,dex.
This is certainly due to the differences between  the model atoms used in the
two investigations.

The lowest value of [Cu/Fe] in our sample is --0.35 for HD\,140283, while
for the most metal-poor star
in our sample, CD\,--38$^\circ$\,245, the Cu to Fe ratio is essentially
solar, like in CS\,31082-001.
From our sample we are not in a position to conclude in a robust way if there is a trend 
of [Cu/Fe] with [Fe/H] or if there is simply a scatter
in [Cu/Fe] abudnances.

If we take our [Cu/Fe] ratios at face value and take the mean of the sample
we find $\left\langle [\rm Cu/Fe] \right\rangle = -0.15$ with a standard
deviation of 0.15\,dex. The weighted mean is $-0.22$\,dex and the 
mean error of unit weight \citep{agekan} is 0.038\,dex. The fact that 
this is smaller than the standard deviation suggests that the value of [Cu/Fe]
is not constant, yet the small size of our sample does not warrant
any more detailed statistical analysis. 
We plan, in the future to apply the NLTE analysis to a much larger
sample of stars for which the Cu resonance lines can be measured.
A larger sample may allow to determine in a robust way the Galactic evolution of Cu.

\section{Conclusions}

Our NLTE model atom allows to derive consistent copper
abundances
from several atomic lines, over a wide range of stellar
parameters, including metallicity. 
In agreement with previous investigations we confirm that 
the deviations from NLTE are larger for lower metallicities, 
yet our NLTE corrections are larger than those that 
have been found by other groups for the lines in common \citet{Shi14,Yan15,Yan16}. 
We have been able to show that the UV \ion{Cu}{i} resonance lines, that are measurable
down to very low metallicities ([Cu/H]$\le -4.2$), can be used as abundance indicators, 
if treated in NLTE. 

The influence of granulation effects on the formation of these lines remains 
an open issue. \citet{Bonifacio10} have shown that LTE computations, based
on 3D hydrodynamical simulations imply large corrections. Yet those computations
were unable to achieve consistency between the UV \ion{Cu}{i} resonant doublet and
the 5105\,\AA\ line. Therefore the issue has to be addressed by performing 3D-NLTE 
computations, that are not currently available. 

The picture of the Galactic evolution of Cu that emerges from our analysis (see Fig.\,\ref{cu_fe})
is very different from what has become familiar from previous LTE analysis
\citep{Mishenina02,Simmerer,Bihain}, and that is essentially
confirmed by our own LTE analysis, in spite of the limited size
of the sample.
Our results provide a picture that differs also  from the NLTE analysis
of \citet{Yan15}. Our small sample is consistent with a mean
constant value of [Cu/Fe]=--0.22 and a scatter of 0.15\,dex, in excess of
the observational error for the [Fe/H] range covered by our
sample. A decrease of [Cu/Fe] from a nearly solar value at [Fe/H]=--1.5
down to --0.35 at [Fe/H]=--2.5, followed by a steep increase to 0.0
cannot be excluded either.
A much larger sample is needed to clarify the situation, it is however clear
that a NLTE treatment, using our model atom, will result in [Cu/Fe] ratios
that are considerably larger than those found by other investigations. 
We can say that, by and large, Cu tracks closely Fe.
This is something that models of Galactic evolution will have to 
take into account.

\section*{Acknowledgements}

We are grateful to Ruth Peterson for giving us
the reduced HST spectra that have been used in this
paper and for useful comments on the manuscript.
SMA is thankful to the GEPI Department and Paris Observatoire administration 
for their hospitality during his visit, and to the Scientific Council of Observatoire 
de Paris  for the financial support. SAK and SMA acknowledge the partial financial 
support from the SCOPES grant No. IZ73Z0-152485. SAK and AVZh are thankful
to the Crimean Council of Ministers for the RFBR grant No. 17-42-92008.




\bibliographystyle{mnras}
\bibliography{cunlte} 

\begin{thebibliography}{}
\makeatletter
\relax
\def\mn@urlcharsother{\let\do\@makeother \do\$\do\&\do\#\do\^\do\_\do\%\do\~}
\def\mn@doi{\begingroup\mn@urlcharsother \@ifnextchar [ {\mn@doi@}
  {\mn@doi@[]}}
\def\mn@doi@[#1]#2{\def\@tempa{#1}\ifx\@tempa\@empty \href
  {http://dx.doi.org/#2} {doi:#2}\else \href {http://dx.doi.org/#2} {#1}\fi
  \endgroup}
\def\mn@eprint#1#2{\mn@eprint@#1:#2::\@nil}
\def\mn@eprint@arXiv#1{\href {http://arxiv.org/abs/#1} {{\tt arXiv:#1}}}
\def\mn@eprint@dblp#1{\href {http://dblp.uni-trier.de/rec/bibtex/#1.xml}
  {dblp:#1}}
\def\mn@eprint@#1:#2:#3:#4\@nil{\def\@tempa {#1}\def\@tempb {#2}\def\@tempc
  {#3}\ifx \@tempc \@empty \let \@tempc \@tempb \let \@tempb \@tempa \fi \ifx
  \@tempb \@empty \def\@tempb {arXiv}\fi \@ifundefined
  {mn@eprint@\@tempb}{\@tempb:\@tempc}{\expandafter \expandafter \csname
  mn@eprint@\@tempb\endcsname \expandafter{\@tempc}}}

\bibitem[\protect\citeauthoryear{{Agekan}}{{Agekan}}{1972}]{agekan}
{Agekan} T.~A.,  1972, {Osnovi teorii osibok dla astronomov i fizikov}.
Izdavateljstvo ``Nauka'' glavna redakcia fizicko-matematceskog literaturi,
  Moskva

\bibitem[\protect\citeauthoryear{{Allen}}{{Allen}}{1973}]{Allen73}
{Allen} C.~W.,  1973, {Astrophysical quantities}.
University of London, Athlone Press

\bibitem[\protect\citeauthoryear{{Bihain}, {Israelian}, {Rebolo}, {Bonifacio}
  \& {Molaro}}{{Bihain} et~al.}{2004}]{Bihain}
{Bihain} G.,  {Israelian} G.,  {Rebolo} R.,  {Bonifacio} P.,   {Molaro} P.,
  2004, \mn@doi [\aap] {10.1051/0004-6361:20035913}, \href
  {http://esoads.eso.org/abs/2004A%26A...423..777B} {423, 777}

\bibitem[\protect\citeauthoryear{{Bisterzo}, {Gallino}, {Pignatari}, {Pompeia},
  {Cunha}  \& {Smith}}{{Bisterzo} et~al.}{2004}]{Bisterzo04}
{Bisterzo} S.,  {Gallino} R.,  {Pignatari} M.,  {Pompeia} L.,  {Cunha} K.,
  {Smith} V.,  2004, \memsai, \href
  {http://cdsads.u-strasbg.fr/abs/2004MmSAI..75..741B} {75, 741}

\bibitem[\protect\citeauthoryear{{Bonifacio}, {Caffau}  \&
  {Ludwig}}{{Bonifacio} et~al.}{2010}]{Bonifacio10}
{Bonifacio} P.,  {Caffau} E.,   {Ludwig} H.-G.,  2010, \mn@doi [\aap]
  {10.1051/0004-6361/200912935}, \href
  {http://esoads.eso.org/abs/2010A%26A...524A..96B} {524, A96}

\bibitem[\protect\citeauthoryear{{Carlsson}}{{Carlsson}}{1986}]{Car86}
{Carlsson} M.,  1986, Uppsala Astronomical Observatory Reports, \href
  {http://adsabs.harvard.edu/abs/1986UppOR..33.....C} {33}

\bibitem[\protect\citeauthoryear{{Carlsson}}{{Carlsson}}{1988}]{1988PhRvA..38.1702C}
{Carlsson} J.,  1988, \mn@doi [\pra] {10.1103/PhysRevA.38.1702}, \href
  {http://esoads.eso.org/abs/1988PhRvA..38.1702C} {38, 1702}

\bibitem[\protect\citeauthoryear{{Castelli} \& {Kurucz}}{{Castelli} \&
  {Kurucz}}{2003}]{CK03}
{Castelli} F.,  {Kurucz} R.~L.,  2003, in {Piskunov} N.,  {Weiss} W.~W.,
  {Gray} D.~F.,  eds,  IAU Symposium Vol. 210, Modelling of Stellar
  Atmospheres. p.~A20 (\mn@eprint {} {astro-ph/0405087})

\bibitem[\protect\citeauthoryear{{Cayrel} et~al.,}{{Cayrel}
  et~al.}{2004}]{FirstStarsV}
{Cayrel} R.,  et~al., 2004, \mn@doi [\aap] {10.1051/0004-6361:20034074}, \href
  {http://cdsads.u-strasbg.fr/abs/2004A%26A...416.1117C} {416, 1117}

\bibitem[\protect\citeauthoryear{{Christlieb}, {Gustafsson}, {Korn}, {Barklem},
  {Beers}, {Bessell}, {Karlsson}  \& {Mizuno-Wiedner}}{{Christlieb}
  et~al.}{2004}]{chris2004}
{Christlieb} N.,  {Gustafsson} B.,  {Korn} A.~J.,  {Barklem} P.~S.,  {Beers}
  T.~C.,  {Bessell} M.~S.,  {Karlsson} T.,   {Mizuno-Wiedner} M.,  2004,
  \mn@doi [\apj] {10.1086/381237}, \href
  {http://cdsads.u-strasbg.fr/abs/2004ApJ...603..708C} {603, 708}

\bibitem[\protect\citeauthoryear{{Cohen}}{{Cohen}}{1980}]{Cohen80}
{Cohen} J.~G.,  1980, \mn@doi [\apj] {10.1086/158412}, \href
  {http://cdsads.u-strasbg.fr/abs/1980ApJ...241..981C} {241, 981}

\bibitem[\protect\citeauthoryear{{Cohen}, {Christlieb}, {McWilliam},
  {Shectman}, {Thompson}, {Melendez}, {Wisotzki}  \& {Reimers}}{{Cohen}
  et~al.}{2008}]{Cohen2008}
{Cohen} J.~G.,  {Christlieb} N.,  {McWilliam} A.,  {Shectman} S.,  {Thompson}
  I.,  {Melendez} J.,  {Wisotzki} L.,   {Reimers} D.,  2008, \mn@doi [\apj]
  {10.1086/523638}, \href {http://cdsads.u-strasbg.fr/abs/2008ApJ...672..320C}
  {672, 320}

\bibitem[\protect\citeauthoryear{{Drawin}}{{Drawin}}{1968}]{Drawin68}
{Drawin} H.-W.,  1968, \mn@doi [Zeitschrift fur Physik] {10.1007/BF01379963},
  \href {http://esoads.eso.org/abs/1968ZPhy..211..404D} {211, 404}

\bibitem[\protect\citeauthoryear{{Drawin}}{{Drawin}}{1969}]{Drawin69}
{Drawin} H.~W.,  1969, \mn@doi [Zeitschrift fur Physik] {10.1007/BF01392775},
  \href {http://esoads.eso.org/abs/1969ZPhy..225..483D} {225, 483}

\bibitem[\protect\citeauthoryear{{Duffau} et~al.,}{{Duffau}
  et~al.}{2017}]{duffau}
{Duffau} S.,  et~al., 2017, preprint, \href
  {http://cdsads.u-strasbg.fr/abs/2017arXiv170402981D} {} (\mn@eprint {arXiv}
  {1704.02981})

\bibitem[\protect\citeauthoryear{{Gerstenberger}, {Latush}  \&
  {Collins}}{{Gerstenberger} et~al.}{1979}]{1979OptCo..31...28G}
{Gerstenberger} D.~C.,  {Latush} E.~L.,   {Collins} G.~J.,  1979, \mn@doi
  [Optics Communications] {10.1016/0030-4018(79)90237-2}, \href
  {http://cdsads.u-strasbg.fr/abs/1979OptCo..31...28G} {31, 28}

\bibitem[\protect\citeauthoryear{{Grevesse}, {Scott}, {Asplund}  \&
  {Sauval}}{{Grevesse} et~al.}{2015}]{2015A&A...573A..27G}
{Grevesse} N.,  {Scott} P.,  {Asplund} M.,   {Sauval} A.~J.,  2015, \mn@doi
  [\aap] {10.1051/0004-6361/201424111}, \href
  {http://cdsads.u-strasbg.fr/abs/2015A%26A...573A..27G} {573, A27}

\bibitem[\protect\citeauthoryear{{Hermann}, {Lasnitschka}, {Schwabe}  \&
  {Spengler}}{{Hermann} et~al.}{1993}]{1993AcSpe..48.1259H}
{Hermann} G.,  {Lasnitschka} G.,  {Schwabe} C.,   {Spengler} D.,  1993, \mn@doi
  [Spectrochimica Acta] {10.1016/0584-8547(93)80110-G}, \href
  {http://cdsads.u-strasbg.fr/abs/1993AcSpe..48.1259H} {48, 1259}

\bibitem[\protect\citeauthoryear{{Kock} \& {Richter}}{{Kock} \&
  {Richter}}{1968}]{1968ZA.....69..180K}
{Kock} M.,  {Richter} J.,  1968, \zap, \href
  {http://esoads.eso.org/abs/1968ZA.....69..180K} {69, 180}

\bibitem[\protect\citeauthoryear{{Korotin}, {Andrievsky}  \& {Luck}}{{Korotin}
  et~al.}{1999}]{Kor99}
{Korotin} S.~A.,  {Andrievsky} S.~M.,   {Luck} R.~E.,  1999, \aap, \href
  {http://adsabs.harvard.edu/abs/1999A%26A...351..168K} {351, 168}

\bibitem[\protect\citeauthoryear{{Kurucz}}{{Kurucz}}{1993}]{1993KurCD..13.....K}
{Kurucz} R.,  1993, ATLAS9 Stellar Atmosphere Programs and 2 km/s grid.~Kurucz
  CD-ROM No.~13.~ Cambridge, Mass.: Smithsonian Astrophysical Observatory,
  1993., \href {http://cdsads.u-strasbg.fr/abs/1993KurCD..13.....K} {13}

\bibitem[\protect\citeauthoryear{{Kurucz}}{{Kurucz}}{2005a}]{2005MSAIS...8...14K}
{Kurucz} R.~L.,  2005a, Memorie della Societa Astronomica Italiana Supplementi,
  \href {http://cdsads.u-strasbg.fr/abs/2005MSAIS...8...14K} {8, 14}

\bibitem[\protect\citeauthoryear{{Kurucz}}{{Kurucz}}{2005b}]{2005MSAIS...8...86K}
{Kurucz} R.~L.,  2005b, Memorie della Societa Astronomica Italiana Supplementi,
  \href {http://cdsads.u-strasbg.fr/abs/2005MSAIS...8...86K} {8, 86}

\bibitem[\protect\citeauthoryear{{Kurucz}}{{Kurucz}}{2011}]{2011CaJPh..89..417K}
{Kurucz} R.~L.,  2011, \mn@doi [Canadian Journal of Physics] {10.1139/p10-104},
  \href {http://cdsads.u-strasbg.fr/abs/2011CaJPh..89..417K} {89, 417}

\bibitem[\protect\citeauthoryear{{Kurucz}}{{Kurucz}}{2014}]{2014dapb.book...63K}
{Kurucz} R.~L.,  2014, {Problems with Atomic and Molecular Data: Including All
  the Lines}.
pp 63--73, \mn@doi{10.1007/978-3-319-06956-2_6}

\bibitem[\protect\citeauthoryear{{Kurucz}, {Furenlid}, {Brault}  \&
  {Testerman}}{{Kurucz} et~al.}{1984}]{1984sfat.book.....K}
{Kurucz} R.~L.,  {Furenlid} I.,  {Brault} J.,   {Testerman} L.,  1984, {Solar
  flux atlas from 296 to 1300 nm}

\bibitem[\protect\citeauthoryear{{Lai}, {Bolte}, {Johnson}, {Lucatello},
  {Heger}  \& {Woosley}}{{Lai} et~al.}{2008}]{Lai08}
{Lai} D.~K.,  {Bolte} M.,  {Johnson} J.~A.,  {Lucatello} S.,  {Heger} A.,
  {Woosley} S.~E.,  2008, \mn@doi [\apj] {10.1086/588811}, \href
  {http://cdsads.u-strasbg.fr/abs/2008ApJ...681.1524L} {681, 1524}

\bibitem[\protect\citeauthoryear{{Liu}, {Gao}, {Zeng}, {Yuan}  \& {Shi}}{{Liu}
  et~al.}{2014}]{Liu2014}
{Liu} Y.~P.,  {Gao} C.,  {Zeng} J.~L.,  {Yuan} J.~M.,   {Shi} J.~R.,  2014,
  \mn@doi [\apjs] {10.1088/0067-0049/211/2/30}, \href
  {http://esoads.eso.org/abs/2014ApJS..211...30L} {211, 30}

\bibitem[\protect\citeauthoryear{{Lodders}, {Palme}  \& {Gail}}{{Lodders}
  et~al.}{2009}]{Lodders}
{Lodders} K.,  {Palme} H.,   {Gail} H.-P.,  2009, \mn@doi [Landolt
  B{\"o}rnstein] {10.1007/978-3-540-88055-4_34}, \href
  {http://cdsads.u-strasbg.fr/abs/2009LanB...4B...44L} {}

\bibitem[\protect\citeauthoryear{{Mishenina}, {Kovtyukh}, {Soubiran},
  {Travaglio}  \& {Busso}}{{Mishenina} et~al.}{2002}]{Mishenina02}
{Mishenina} T.~V.,  {Kovtyukh} V.~V.,  {Soubiran} C.,  {Travaglio} C.,
  {Busso} M.,  2002, \mn@doi [\aap] {10.1051/0004-6361:20021399}, \href
  {http://cdsads.u-strasbg.fr/abs/2002A%26A...396..189M} {396, 189}

\bibitem[\protect\citeauthoryear{{Nissen} \& {Schuster}}{{Nissen} \&
  {Schuster}}{2011}]{2011A&A...530A..15N}
{Nissen} P.~E.,  {Schuster} W.~J.,  2011, \mn@doi [\aap]
  {10.1051/0004-6361/201116619}, \href
  {http://cdsads.u-strasbg.fr/abs/2011A%26A...530A..15N} {530, A15}

\bibitem[\protect\citeauthoryear{{Roederer} et~al.,}{{Roederer}
  et~al.}{2012}]{2012ApJS..203...27R}
{Roederer} I.~U.,  et~al., 2012, \mn@doi [\apjs] {10.1088/0067-0049/203/2/27},
  \href {http://cdsads.u-strasbg.fr/abs/2012ApJS..203...27R} {203, 27}

\bibitem[\protect\citeauthoryear{{Roederer} et~al.,}{{Roederer}
  et~al.}{2014}]{2014ApJ...791...32R}
{Roederer} I.~U.,  et~al., 2014, \mn@doi [\apj] {10.1088/0004-637X/791/1/32},
  \href {http://cdsads.u-strasbg.fr/abs/2014ApJ...791...32R} {791, 32}

\bibitem[\protect\citeauthoryear{{Ryabchikova}, {Piskunov}, {Kurucz},
  {Stempels}, {Heiter}, {Pakhomov}  \& {Barklem}}{{Ryabchikova}
  et~al.}{2015}]{2015PhyS...90e4005R}
{Ryabchikova} T.,  {Piskunov} N.,  {Kurucz} R.~L.,  {Stempels} H.~C.,  {Heiter}
  U.,  {Pakhomov} Y.,   {Barklem} P.~S.,  2015, \mn@doi [\physscr]
  {10.1088/0031-8949/90/5/054005}, \href
  {http://adsabs.harvard.edu/abs/2015PhyS...90e4005R} {90, 054005}

\bibitem[\protect\citeauthoryear{{Sbordone}}{{Sbordone}}{2005}]{2005MSAIS...8...61S}
{Sbordone} L.,  2005, Memorie della Societa Astronomica Italiana Supplementi,
  \href {http://cdsads.u-strasbg.fr/abs/2005MSAIS...8...61S} {8, 61}

\bibitem[\protect\citeauthoryear{{Sbordone}, {Bonifacio}, {Castelli}  \&
  {Kurucz}}{{Sbordone} et~al.}{2004}]{2004MSAIS...5...93S}
{Sbordone} L.,  {Bonifacio} P.,  {Castelli} F.,   {Kurucz} R.~L.,  2004,
  Memorie della Societa Astronomica Italiana Supplementi, \href
  {http://cdsads.u-strasbg.fr/abs/2004MSAIS...5...93S} {5, 93}

\bibitem[\protect\citeauthoryear{{Sbordone}, {Caffau}, {Bonifacio}  \&
  {Duffau}}{{Sbordone} et~al.}{2014}]{mygisfos}
{Sbordone} L.,  {Caffau} E.,  {Bonifacio} P.,   {Duffau} S.,  2014, \mn@doi
  [\aap] {10.1051/0004-6361/201322430}, \href
  {http://cdsads.u-strasbg.fr/abs/2014A%26A...564A.109S} {564, A109}

\bibitem[\protect\citeauthoryear{{Seaton}}{{Seaton}}{1962}]{Seaton62}
{Seaton} M.~J.,  1962, in Bates D.~R.,  ed., Atomic and Molecular Processes.
  New York: Academic Press

\bibitem[\protect\citeauthoryear{{Shi}, {Gehren}, {Zeng}, {Mashonkina}  \&
  {Zhao}}{{Shi} et~al.}{2014}]{Shi14}
{Shi} J.~R.,  {Gehren} T.,  {Zeng} J.~L.,  {Mashonkina} L.,   {Zhao} G.,  2014,
  \mn@doi [\apj] {10.1088/0004-637X/782/2/80}, \href
  {http://cdsads.u-strasbg.fr/abs/2014ApJ...782...80S} {782, 80}

\bibitem[\protect\citeauthoryear{{Simmerer}, {Sneden}, {Ivans}, {Kraft},
  {Shetrone}  \& {Smith}}{{Simmerer} et~al.}{2003}]{Simmerer}
{Simmerer} J.,  {Sneden} C.,  {Ivans} I.~I.,  {Kraft} R.~P.,  {Shetrone} M.~D.,
    {Smith} V.~V.,  2003, \mn@doi [\aj] {10.1086/373926}, \href
  {http://cdsads.u-strasbg.fr/abs/2003AJ....125.2018S} {125, 2018}

\bibitem[\protect\citeauthoryear{{Siqueira-Mello}, {Andrievsky}, {Barbuy},
  {Spite}, {Spite}  \& {Korotin}}{{Siqueira-Mello}
  et~al.}{2015}]{2015A&A...584A..86S}
{Siqueira-Mello} C.,  {Andrievsky} S.~M.,  {Barbuy} B.,  {Spite} M.,  {Spite}
  F.,   {Korotin} S.~A.,  2015, \mn@doi [\aap] {10.1051/0004-6361/201526695},
  \href {http://adsabs.harvard.edu/abs/2015A%26A...584A..86S} {584, A86}

\bibitem[\protect\citeauthoryear{{Sitnova} et~al.,}{{Sitnova}
  et~al.}{2015}]{Sitnova}
{Sitnova} T.,  et~al., 2015, \mn@doi [\apj] {10.1088/0004-637X/808/2/148},
  \href {http://adsabs.harvard.edu/abs/2015ApJ...808..148S} {808, 148}

\bibitem[\protect\citeauthoryear{{Smiljanic} et~al.,}{{Smiljanic}
  et~al.}{2014}]{2014A&A...570A.122S}
{Smiljanic} R.,  et~al., 2014, \mn@doi [\aap] {10.1051/0004-6361/201423937},
  \href {http://cdsads.u-strasbg.fr/abs/2014A%26A...570A.122S} {570, A122}

\bibitem[\protect\citeauthoryear{{Sneden}, {Gratton}  \& {Crocker}}{{Sneden}
  et~al.}{1991}]{Sneden91}
{Sneden} C.,  {Gratton} R.~G.,   {Crocker} D.~A.,  1991, \aap, \href
  {http://cdsads.u-strasbg.fr/abs/1991A%26A...246..354S} {246, 354}

\bibitem[\protect\citeauthoryear{{Spite}, {Peterson}, {Gallagher}, {Barbuy}  \&
  {Spite}}{{Spite} et~al.}{2017}]{Spite17}
{Spite} M.,  {Peterson} R.~C.,  {Gallagher} A.~J.,  {Barbuy} B.,   {Spite} F.,
  2017, \mn@doi [\aap] {10.1051/0004-6361/201630058}, \href
  {http://cdsads.u-strasbg.fr/abs/2017A%26A...600A..26S} {600, A26}

\bibitem[\protect\citeauthoryear{{Steenbock} \& {Holweger}}{{Steenbock} \&
  {Holweger}}{1984}]{SH84}
{Steenbock} W.,  {Holweger} H.,  1984, \aap, \href
  {http://esoads.eso.org/abs/1984A%26A...130..319S} {130, 319}

\bibitem[\protect\citeauthoryear{{Sugar} \& {Musgrove}}{{Sugar} \&
  {Musgrove}}{1990}]{1990JPCRD..19..527S}
{Sugar} J.,  {Musgrove} A.,  1990, \mn@doi [Journal of Physical and Chemical
  Reference Data] {10.1063/1.555855}, \href
  {http://esoads.eso.org/abs/1990JPCRD..19..527S} {19, 527}

\bibitem[\protect\citeauthoryear{{Summers}}{{Summers}}{2004}]{ADAS}
{Summers} H.~P.,  2004, The ADAS User Manual, version 2.6,
  \url{http://www.adas.ac.uk}

\bibitem[\protect\citeauthoryear{{Tsymbal}}{{Tsymbal}}{1996}]{Tsym96}
{Tsymbal} V.,  1996, in {Adelman} S.~J.,  {Kupka} F.,   {Weiss} W.~W.,  eds,
  Astronomical Society of the Pacific Conference Series Vol. 108, M.A.S.S.,
  Model Atmospheres and Spectrum Synthesis. p.~198

\bibitem[\protect\citeauthoryear{{Vernazza}, {Avrett}  \& {Loeser}}{{Vernazza}
  et~al.}{1981}]{Vernazza}
{Vernazza} J.~E.,  {Avrett} E.~H.,   {Loeser} R.,  1981, \mn@doi [\apjs]
  {10.1086/190731}, \href {http://cdsads.u-strasbg.fr/abs/1981ApJS...45..635V}
  {45, 635}

\bibitem[\protect\citeauthoryear{{Yan}, {Shi}  \& {Zhao}}{{Yan}
  et~al.}{2015}]{Yan15}
{Yan} H.~L.,  {Shi} J.~R.,   {Zhao} G.,  2015, \mn@doi [\apj]
  {10.1088/0004-637X/802/1/36}, \href
  {http://cdsads.u-strasbg.fr/abs/2015ApJ...802...36Y} {802, 36}

\bibitem[\protect\citeauthoryear{{Yan}, {Shi}, {Nissen}  \& {Zhao}}{{Yan}
  et~al.}{2016}]{Yan16}
{Yan} H.~L.,  {Shi} J.~R.,  {Nissen} P.~E.,   {Zhao} G.,  2016, \mn@doi [\aap]
  {10.1051/0004-6361/201527491}, \href
  {http://cdsads.u-strasbg.fr/abs/2016A%26A...585A.102Y} {585, A102}

\bibitem[\protect\citeauthoryear{{Yong}, {Mel{\'e}ndez}, {Cunha}, {Karakas},
  {Norris}  \& {Smith}}{{Yong} et~al.}{2008}]{Yong}
{Yong} D.,  {Mel{\'e}ndez} J.,  {Cunha} K.,  {Karakas} A.~I.,  {Norris} J.~E.,
   {Smith} V.~V.,  2008, \mn@doi [\apj] {10.1086/592229}, \href
  {http://cdsads.u-strasbg.fr/abs/2008ApJ...689.1020Y} {689, 1020}

\bibitem[\protect\citeauthoryear{{van Regemorter}}{{van
  Regemorter}}{1962}]{vanRegemorter}
{van Regemorter} H.,  1962, \mn@doi [\apj] {10.1086/147445}, \href
  {http://esoads.eso.org/abs/1962ApJ...136..906V} {136, 906}

\makeatother
\end{thebibliography}








\bsp	
\label{lastpage}
\end{document}